\documentclass[10pt,journal,compsoc]{IEEEtran}
\usepackage{graphicx}
\usepackage{amsmath}
\usepackage{algorithmicx}
\usepackage[ruled]{algorithm}
\usepackage{algpseudocode}
\usepackage{multicol}
\usepackage{xcolor}
\usepackage{times}
\usepackage{helvet}
\usepackage{courier}
\usepackage{graphicx}
\usepackage{paralist}
\usepackage{helvet} 
\usepackage{courier}  
\usepackage{multirow}
\usepackage{marvosym}
\usepackage{fontawesome}
\usepackage{newfloat}
\usepackage{listings}
\usepackage{caption}

\ifCLASSOPTIONcompsoc
  \usepackage[nocompress]{cite}
\else
  \usepackage{cite}
\fi
\ifCLASSINFOpdf
\else
\fi
\begin{document}
%
\title{Quantifying the Academic Quality of Children's Videos using Machine Comprehension}
%
%
%
%

\author{ Sumeet Kumar, Mallikarjuna T., 
Ashiqur Khudabukhsh\\  
sumeet.kumar@sv.cmu.edu, mallikarjuna\_tupakula@isb.edu, axkvse@rit.edu\\
}

\IEEEtitleabstractindextext{%
\begin{abstract}

YouTube Kids (YTK) is one of the most popular kids' applications used by millions of kids daily. However, various studies have highlighted concerns about the videos on the platform, like the over-presence of entertaining and commercial content. YouTube recently proposed high-quality guidelines that include `promoting learning' and proposed to use it in ranking channels. However, the concept of learning is multi-faceted, and it can be difficult to define and measure in the context of online videos. This research focuses on learning in terms of what's taught in schools and proposes a way to measure the academic quality of children's videos. Using a new dataset of questions and answers from children's videos, we first show that a Reading Comprehension (RC) model can estimate academic learning. Then, using a large dataset of middle school textbook questions on diverse topics, we quantify the academic quality of top channels as the number of children's textbook questions that an RC model can correctly answer. By analyzing over 80,000 videos posted on the top 100 channels, we present the first thorough analysis of the academic quality of channels on YTK. 

\end{abstract}

}

\maketitle

\IEEEdisplaynontitleabstractindextext

%
\IEEEpeerreviewmaketitle

\IEEEraisesectionheading{\section{Introduction}\label{sec:introduction}}

YouTube Kids (YTK) is a widely used video application and web platform among children attracting millions of daily users worldwide~\cite{radesky2020young}. The platform is designed to provide a safe and kid-friendly environment for children to explore and discover online content. However, studies have raised concerns about the types of videos available on the platform highlighting the over-presence of entertaining and commercial content~\cite{jaakkola2020vernacularized}, and the potential exposure of children to inappropriate or harmful content~\cite{papadamou2020disturbed}. In the past, YouTube video creators could record just about whatever they wanted, and the content could not be removed unless there were any violations. While it is important to screen videos for inappropriate content, the absence of such content does not automatically imply that the videos are good for kids. For instance, ``toy opening" videos, which are one of the most viewed genres on YTK\footnote{https://www.bbc.com/news/uk-england-beds-bucks-herts-49975644}, might not contain anything inappropriate~\cite{nicoll2018mimetic}, yet they may not be considered of high quality~\cite{craig2017toy}. Beyond ensuring the content is safe and appropriate, it is also important that the content is educational and informative.

Research has shown that the type of media children consumes can have a significant impact on their cognitive and socio-emotional development~\cite{rasmussen2019promoting}. Ensuring that children are exposed to high-quality academic content that can support their learning and development is thus important.  With this goal, YouTube has proposed high-quality guidelines that include promoting learning and inspiring curiosity and has proposed to use them in their ranking metrics\footnote{https://support.google.com/youtube/answer/10774223?hl=en-GB}. However, measuring video quality and enforcing them on a platform where thousands of new videos get uploaded every hour is an uphill task. The absence of a metric to judge the quality of videos and inform the viewers about it often leaves important stakeholders (like parents) concerned about the impact of video watching on their children~\cite{evans2018parenting}.

\begin{figure}[t]
    \centering
    \includegraphics[width=0.42\textwidth]{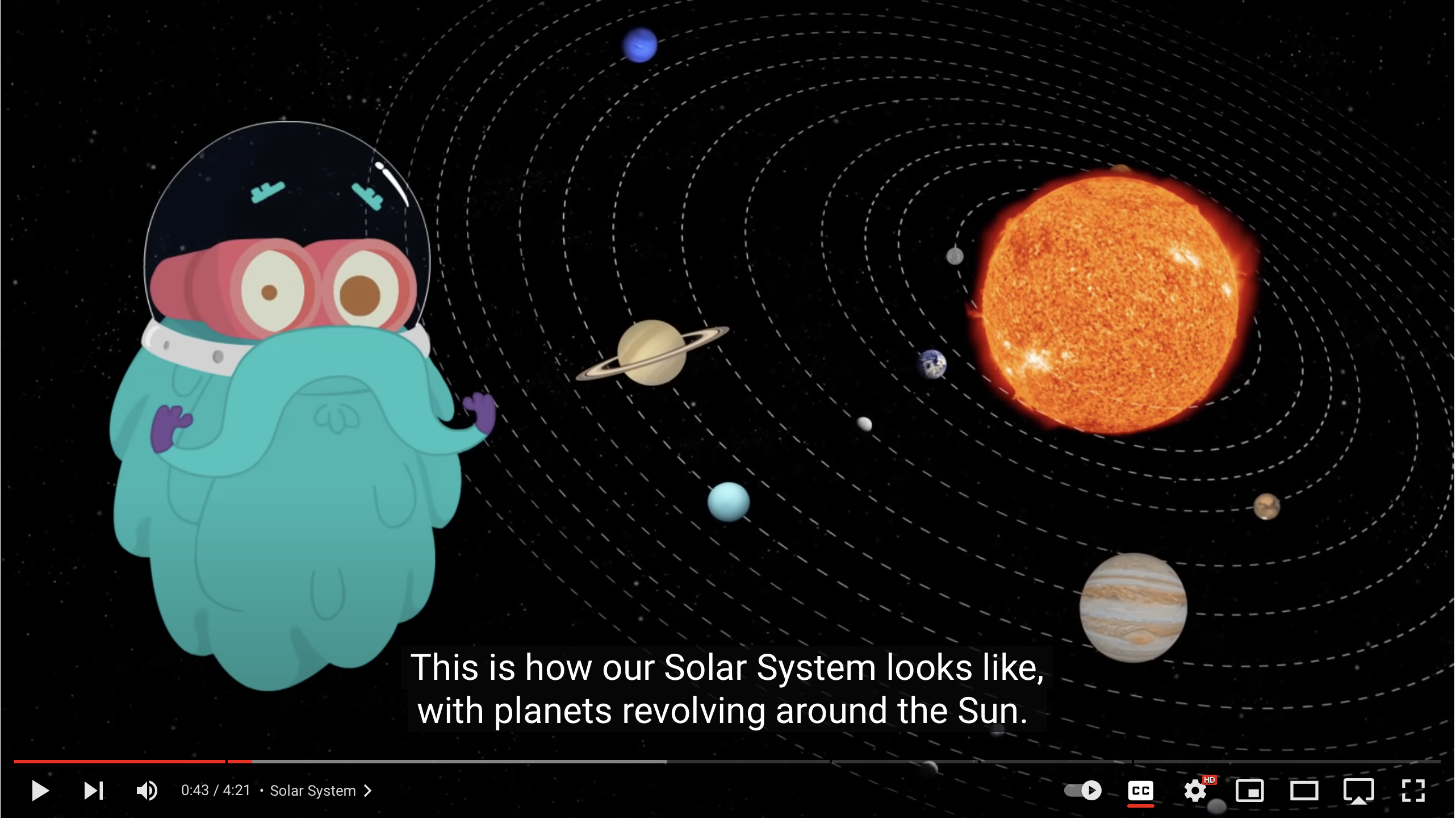}
    \caption[RC Model]{A video-frame from a \textit{YouTube Kids} video explaining the Solar System. Our paper attempts to quantify the academic quality of videos on the basis of visual and language content.}
    \label{fig:earth_is_round}
\end{figure}

What is \textit{quality}? and how can one measure the \textit{quality of a video}? Oxford dictionary defines quality as \textit{`the standard of something when it is compared to other things like it; how good or bad something is'}. To measure learning, we consider what is taught in schools, and propose a way to measure the academic quality of children’s videos by comparing information in videos (e.g., see Fig. \ref{fig:earth_is_round}) to information in standard children's textbooks. While entertaining videos are also higher in entertainment quality, this research focuses on what the young viewers learn from a video vis-à-vis school textbook content, and we call this metric the academic quality. 

To measure academic quality, we propose to use children's textbooks. Children's textbooks are vetted to be informative with high learning outcomes and, therefore, would work as a good baseline to assess the quality. Moreover, they often come with questions and answers that can further be used to validate learning. Therefore, we propose using question-answers from textbooks targeting children to find if these questions could be correctly answered using videos. This method is similar to the process of conducting reading comprehension tests for students. Just as a student's understanding of a passage is measured by their ability to answer questions based on the passage, we measure the academic quality of a video by estimating the number of questions (from children's textbooks) that can be answered by viewing the video.

While viewing a video and answering questions to estimate the academic quality of videos is a reasonable approach, manually, it is infeasible given the scale at which YTK operates. Instead, we want an automated approach that mimics the process. To automate the process of a student reading a paragraph to answer questions in tests, we propose to use a \textit{reading comprehension} (RC) model (as a proxy for a student) that uses information in videos to answer questions relevant to children. For example, as shown in  Fig. \ref{fig:earth_is_round}, by using the information provided in the video, an RC model would be able to answer questions such as ``Do planets revolve around the Sun?". 

Previous research has demonstrated the potential of using RC models for question answering\cite{kembhavi2017you} but assessing the academic quality of videos using RC models brings many challenges. For example, many videos could be long, and most RC models cannot handle long comprehension text. We enhance the RC model by incorporating a global and sliding window attention mechanism to tackle lengthy videos. Besides length, the other concern with video data is the presence of multiple mediums of information, including visuals and audio. In prior research, audio information was transcribed into language and used for question-answering; however, visuals were difficult to use. Our proposed approach also uses visuals by extracting visual information from video frames.


Broadly speaking, our research provides an approach to identify potentially low-academic-quality content on YTK. We develop a method that can facilitate the evaluation of the academic quality of videos on the platform, which in turn, ultimately helps parents and video hosting platforms make more informed decisions about the content that children consume online. Though we focus on English-language science questions, our approach can be extended to measure the academic quality of videos in other languages and subjects and would be helpful to  educators who develop academic content for children. Additionally, our approach can be used to evaluate the effectiveness of the guidelines that YouTube has put in place to promote learning  in children.

To summarize, we estimate the academic quality of videos using a benchmark derived from the number of questions answered correctly by an RC model sourced from middle-school textbooks. Our evaluation only reflects academic quality relative to the middle-school curriculum, and we recognize the limitations of this approach. Nevertheless, we view our work as a crucial advancement toward addressing a significant societal issue. The proposed model leverages both video transcripts and frames as inputs to make predictions about the overall academic quality of videos.

\textbf{Contributions:} Our contributions are the following.
\begin{compactenum}
\item\emph{Method:} 
We propose an automated approach to estimate the educational quality of videos. Our novel approach uses an extension of the machine comprehension framework to multi-modal data for estimating academic quality.
\item \emph{Social:} 
To the best of our knowledge, our work is the first to estimate the academic quality of large-scale kids' videos on YTK. The quality of videos on YTK is one of the important concerns frequently raised by parents, and therefore, this research has great social significance. 
\item \emph{Resource:} 
We share a novel dataset of over 80,000 publicly available kids' video transcripts and salient frames from the 100 top YTK channels. Beyond our current task of estimating academic quality, we believe this dataset will be useful for several other research questions.
\end{compactenum}
  
The paper is organized as follows. We summarize prior work (Sec. \ref{sec:work}) highlighting how our work covers an important gap in analyzing educational videos. We then discuss our methodology (Sec. \ref{sec:methodology}) where in we combine the state-of-the-art document retrieval and machine comprehension models to answer multiple choice questions using multi-modal video data. We describe the datasets used in the experiments in Sec. \ref{sec:dataset}. We then present the results in three separate sub-sections: 1) evaluating RC models on kids' videos data, 2) quantifying the academic quality of videos using \texttt{TQA} dataset, and 3) comparing top YTK channels in terms of academic quality, followed by a discussion of the results in Sec. \ref{sec:exp_results}. Finally, we conclude and propose directions for future work (Sec. \ref{sec:colclusionAndFutureWork}). Supplementary information (SI) has additional details on datasets and the models used in the paper. Source code to reproduce the experimental results will be made available on Github\footnote{https://github.com/sumeetkr/KidsVideosEducationalValue}.

\section{Related Work}\label{sec:work}
In this section, we first discuss recent research highlighting concerns with YTK followed by a literature review of `Reading Comprehension', and `Video Retrieval', the two important threads of research that study analyzing and retrieving videos. We then elaborate on how our approach creates a new direction by combining reading comprehension and video retrieval to quantify video quality.

\subsection{Concerns with YouTube Kids Videos}
Researchers have studied videos on YTK lately, highlighting multiple concerns (\cite{craig2017toy,araujo2017characterizing,neumann2020evaluating}). For example, \cite{craig2017toy} explored toy-unboxing videos, a widely popular genre on YTK, that involves opening, assembling, and demonstrating children’s toys. The authors show how such videos are in the structural and material interests of the social media entertainment industry, raising important regulatory questions. \cite{evans2018parenting} studies influencer unboxing videos to study if they appropriately disclose sponsorship and how  parents understand and respond to sponsored child content, raising questions about such videos' quality. 

The quality of children's content has been studied outside academia as well. For instance, a blog post titled ``Something is wrong on the Internet''\footnote{https://medium.com/@jamesbridle/something-is-wrong-on-the-internet-c39c471271d2} discusses the quality concerns of YTK videos. An organization named Truth in Advertising\footnote{truthinadvertising.org} (TINA)  conducted a detailed review of every video published on the Ryan Toys Review (RTR) channel, one of the top channels on YTK, between January 1 and July 31, 2019.  The investigation revealed that ``92 percent of videos promote at least one product or television/YouTube program that is inappropriate for – and targeted at – children under five'' \footnote{source: https://www.truthinadvertising.org/ryan-toysreview-database/}. The investigation led to FTC's complaint against RTR on August 29, 2019, highlighting the issues with a top channel on YTK (\cite{httpswww47:online}). Unlike existing  efforts highlighting concerns with kids' videos,  we quantify the academic quality of videos to study the platform further.


\subsection{Reading Comprehension}

\begin{figure*}[htb!]
    \centering
    \includegraphics[width=0.8\textwidth]{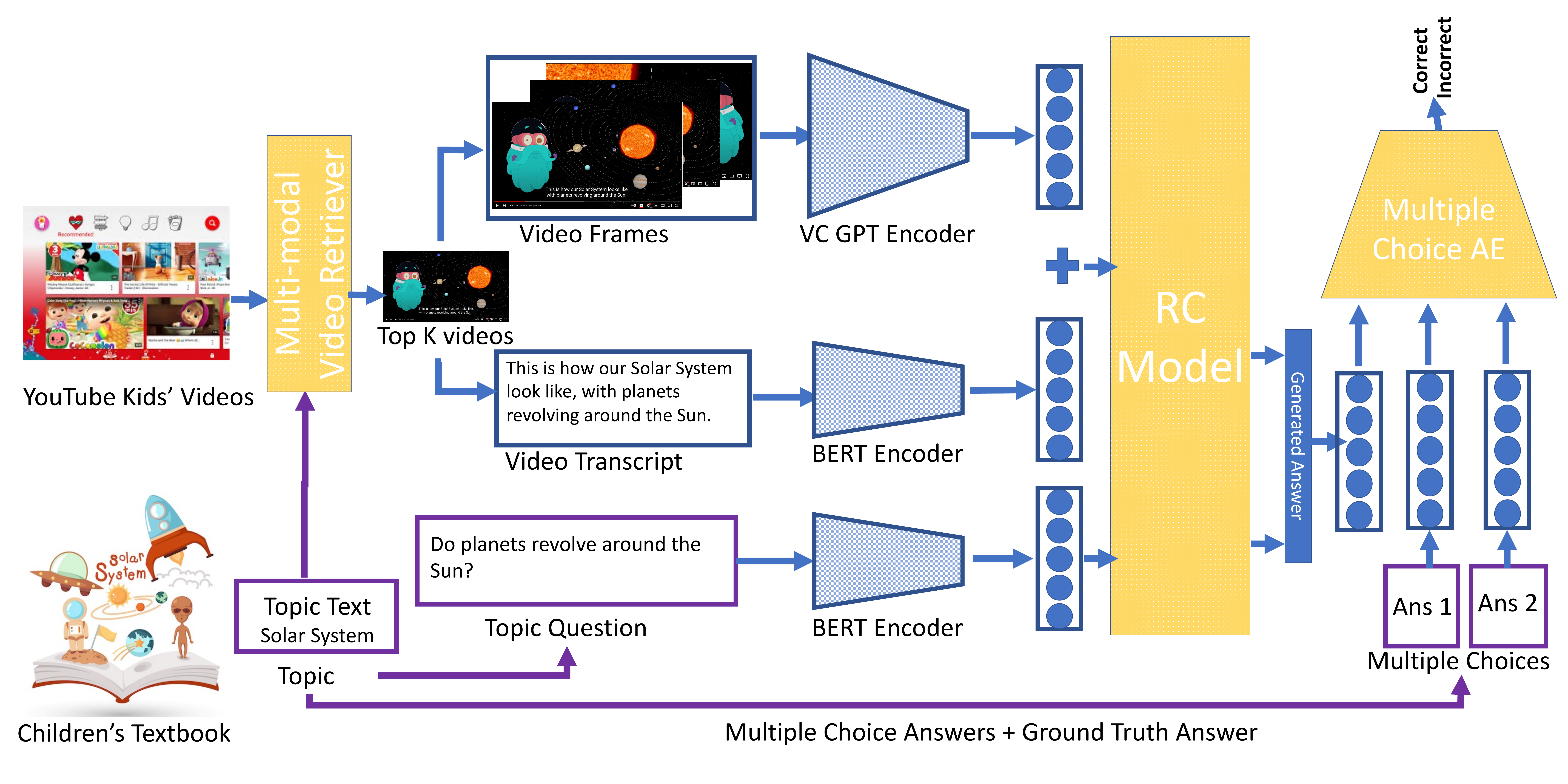}
    \caption[RC Model]{Proposed approach to estimate the academic quality of videos vis-a-vis questions and answers in children's textbooks. The proposed approach combines a multi-modal video retriever, a reading comprehension (RC) model, and an answer extraction (Multiple Choice AE) model. }
    \label{fig:flow}
\end{figure*}

Reading comprehension (RC) tasks have received significant attention in recent literature \cite{song2020evidence,yu2021multi}. An RC model answers questions based on the context information from RC datasets and is evaluated on generated answers. Yu \textit{et al.} \cite{yu2018qanet} proposed a Q\&A architecture named QANet that employs convolution and self-attention to achieve fast training and inference. Yan \textit{et al.} \cite{yan2019deep} proposed a deep cascade learning model that provides a more precise answer extraction in machine reading comprehension. Zhang \textit{et al.} \cite{zhang2020sg} introduced the syntactic dependency of interest (SDOI) design that incorporates explicit syntactic constraints into the attention mechanism to guide text modeling with a syntax-guided self-attention mechanism.

Training datasets, that play a crucial role in RC training, typically consist of a context and a set of questions and answers based on the context. Popular datasets include (1) \texttt{RACE} \cite{lai2017race}, consisting of 28,000 questions used in English exams for Chinese students aged between 12 to 18; the \texttt{WIKIQA} dataset \cite{yang2015wikiqa} designed for open domain question answering; and the \texttt{HOTPO\texttt{TQA}} dataset \cite{yang2018hotpotqa}, a more challenging QA dataset with 113k Wikipedia-based question-answer pairs requiring complex reasoning. Most recently, \cite{kwiatkowski2019natural} presented a new benchmark for natural question-answering research, consisting of queries from the Google search engine.

Visual question answering is a recent advancement that has introduced image-based QA datasets, such as the \texttt{GQA} dataset \cite{hudson2019gqa} for visual reasoning and compositional question answering. This area of research has been extended to video-based question answering \cite{zhao2017video,xu2017video,zhang2019frame}, which presents greater challenges dealing with  multiple frames across time. To address these challenges, multi-modal machine comprehension (M3C) has been introduced \cite{kembhavi2017you}, which involves answering questions based on a combination of text, diagrams, and images. Furthermore, the NarrativeQA reading comprehension challenge was introduced \cite{kovcisky2018narrativeqa}, emphasizing the importance of understanding the narrative rather than relying on shallow pattern matching.


Our work extends this topic by utilizing RC models for quantifying the academic value of videos. Our new dataset consists of videos comprised of video transcripts and frames. Given the length of videos, recent models trained on typical datasets may not perform well on long transcripts. To address this, we have explored the promising work of the Dense Passage Retrieval method \cite{karpukhin2020dense} for open-domain question answering. The Retriever-Reader network proposed by the authors is capable of finding the top passages relevant to the question. We have adopted this approach to work with video transcripts and video frames, with promising results.

\subsection{Video Retrieval}
Video retrieval is one of the fundamental tasks in knowledge discovery and has a rich literature (e.g., see \cite{zhang1997integrated,aslandogan1999techniques,yilmaz2019applying,singh2021end}). 
Recent approaches to video retrieval utilize visual transformers \cite{fan2021multiscale}. For cross-media retrieval efficiency, Zhang \textit{et al.} \cite{zhang2021dah} proposed discrete asymmetric hashing and Wang \textit{et al.} \cite{wang2021similar} proposed sports play retrieval based on deep reinforcement learning with similarity. Gabeur \textit{et al.} \cite{gabeur2020multi} used cross-modal cues for caption-to-video retrieval by encoding different modalities in videos along with temporal information using multi-modal transformer models, resulting in state-of-the-art results in video retrieval.


In the past, information retrieval research focused on retrieving relevant passages given a context, but videos present information in multiple modalities, making them challenging to retrieve, especially when visual information is involved. Furthermore, even if a video is contextually relevant, it may not always provide the correct answers to contextual questions. Unlike prior work, our approach goes beyond retrieval and includes question-answering based on factual questions from children's textbooks. We propose a reading comprehension system that augments a multi-modal video retrieval system to assess the academic quality of videos, an area that has not been explored before.

To summarize, prior research has explored the YTK platform and found convincing evidence that the platform needs additional scrutiny. In particular, the concern about the quality of videos has emerged as the top concern. To the best of our knowledge, we are the first to propose an approach to quantify the academic aspect of kids' videos. Through our work on quantifying academic quality, we aim to broadly support platforms, parents, and policy forums in monitoring kids content online.

\section{Methodology}\label{sec:methodology}

To assess the academic quality of videos for children, we propose a new approach that combines  video retrieval and reading comprehension (RC). We recognize that directly applying RC models to a large number of questions and videos would result in a slow and inefficient process. Therefore, we first use video retrieval to quickly identify a few top relevant videos for a given topic (from children's textbooks). Given a video is relevant to a topic does not necessarily mean it has information to answer the topic questions correctly. Therefore, we employ RC models to answer questions (and find videos that are correctly able to answer questions) based on these relevant videos. This approach allows to save time and resources, as only the most relevant videos are subjected to the slower and more detailed RC analysis. Our proposed approach is visualized in Fig. \ref{fig:flow}. 





\subsection{Problem Formulation}
Given a lesson topic $c$ (from a children's textbook)  such as ``the solar system'' and factoid question $q$ (such as ``What do planets revolve around?''), an RC system is expected to pick the right answer (``Sun'') among the multiple-choice options, defined as $A=\{ans_1, ans_2, ans_3, ans_4\}$. SI (Supplementary Information) on \texttt{TQA} dataset provides more details on how lessons, topics, and questions are extracted from children's textbooks and arranged in the dataset. 

Our video dataset is comprised of $N$ videos $\{V_1, V_2, ...., V_N\}$,  and we first pre-process the video transcripts and video frames to derive encodings for each video. Then, for every topic, the multi-modal retrieval model picks the top $k$ relevant videos. The RC model then uses video encodings to answer questions from the children's textbooks on that topic. Finally, using the generated answer, a third model extracts the most appropriate choice out of multiple-choice answer options $A$ for the questions. The academic quality of a video can be quantified based on the number of questions that get answered correctly using the video's data.

To summarize, the proposed approach of picking the correct answer from multiple-choice options given a lesson topic and related questions from a children's textbook involves three models:
\begin{enumerate}
    \item Multi-modal video retriever model: Retrieving the top $k$ relevant videos from a dataset of $N$ videos $\{V_1, V_2, ...., V_N\}$ for every topic $c$.
    \item Reading comprehension model: Identifying a span of text from the linguistic representation of videos as the textual answer to a question.
    \item Multiple-choice answer extraction model: Extracting the most appropriate answer from the set of multiple-choice options $\{ans_1, ans_2, ans_3, ans_4\}$  based on the textual answer.    
\end{enumerate}

We discuss these models next.

\subsection{Multi-modal Video Retriever Model}
A \textit{retriever model} finds the relevant documents within the corpus of a large set of documents for a given a topic. A document retriever model uses an encoder for documents and another encoder for topic text to map them to a vector. For the problem at hand, given topic text, we need to retrieve the top $k$ videos that are most relevant to the topic.


 For finding the top \emph{k} nearest documents for a topic, \cite{karpukhin2020dense} used a dot product similarity in the encoding space. We build on \cite{karpukhin2020dense}, a state-of-the-art method for passage retrieval, and adapt it to the unique challenges of multi-modal video data. Videos could be viewed as a combination of audio and video frames. We obtain video transcripts, a textual encoding of audio data using YouTube API. We also obtain video captions for video frames of videos using a state-of-the-art deep learning model that generates video frame captions. Both video transcripts and captions are then used for video retrieval, which we describe in detail in this section. 

\begin{figure}[t]
    \centering
    \includegraphics[width=0.47\textwidth]{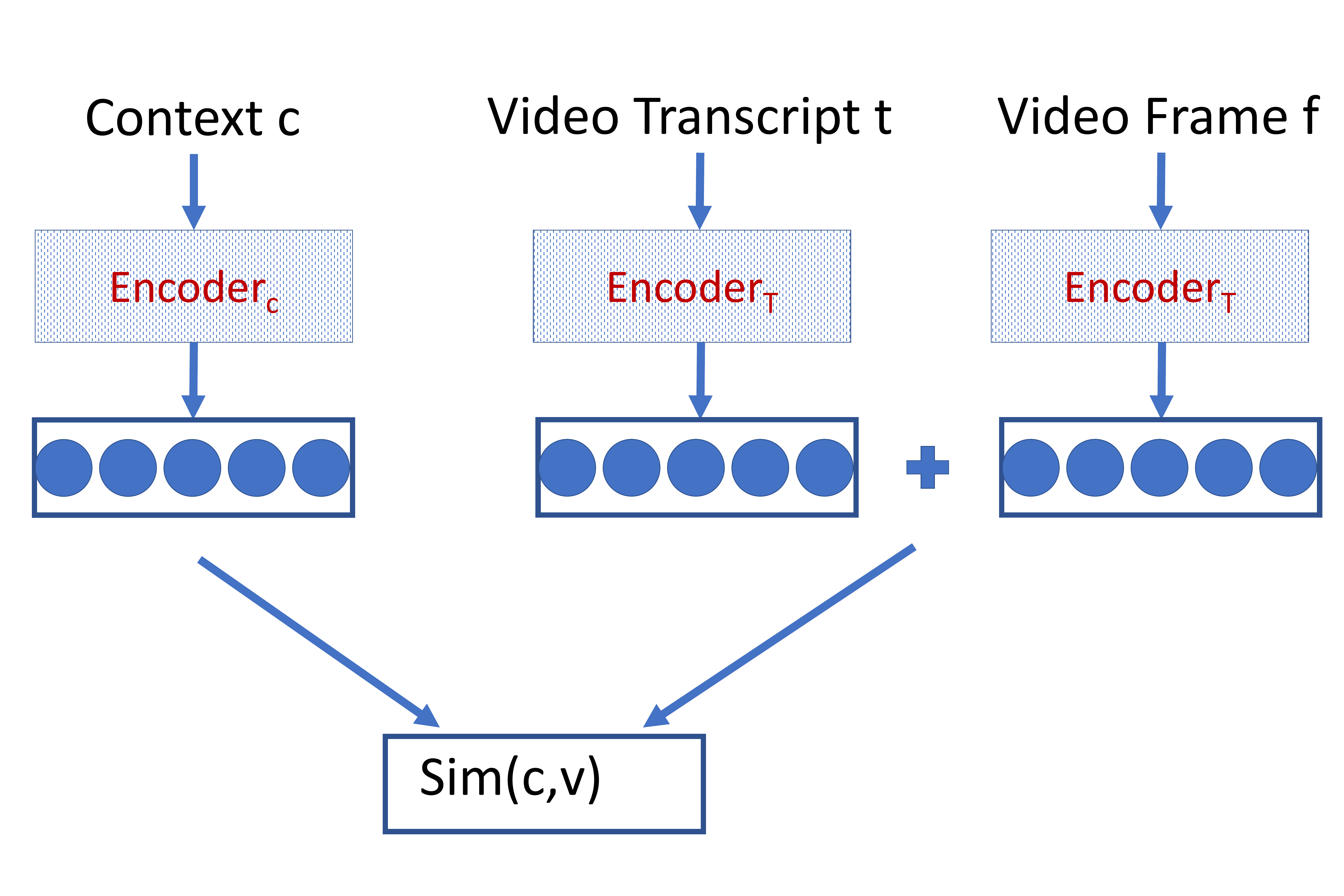}
    \caption[RC Model]{Multi-modal video ranking model, + indicates concatenation of video transcript embeddings and video frame encodings.}
    \label{fig:transcript-ranking}
\end{figure}

As the video transcript is represented as text, it can be directly encoded by a text encoder $E_T$. Similarly, topic text can also be directly encoded $E_C$. 

\begin{equation*}
    E_T(t) = BERT_T(t)
\end{equation*}

\begin{equation*}
    E_C(c) = BERT_C(c)
\end{equation*}

In the above equation, $c$ is the topic (also called context in prior work), and $t$ is a video transcript.  Encoders $E_C$ and $E_T$ are both BERT\cite{devlin2018bert} encoders that take a string (and optionally another one) as arguments and return vectors corresponding to the input tokens. The two BERT encoders ($BERT_C$ and $BERT_T$) do not share weights, which allows them to learn unique aspects of topics description (obtained from textbooks) and videos. 

Video frames need to be first converted to textual representation and then concatenated to form a single text called video captions.

\begin{equation*}
    E_T(f) = BERT_T\Big(\bigoplus_i VCGPT(f_i)\Big)
    \label{eqn:sum_bert}
\end{equation*}

where $\bigoplus$ is the concatenation operation,  $E_T(f)$ is based on Visual Conditioned GPT (VC-GPT)\cite{luo2022vc}, and takes $i^{th}$ frame ($f_i$) of a video as input. VC-GPT uses a pre-trained visual encoder (CLIP-ViT), GPT2 language decoder, and a cross-modal fusion mechanism, leading to state-of-the-art performance on many images captioning datasets \cite{luo2022vc}.

We calculate the similarity between a topic and a video using the dot product of the concatenated outputs of the BERT encoders for the transcript and the VC-GPT encoder for the video frames, as shown in Fig. \ref{fig:transcript-ranking}. For similarity calculation, we are interested in retrieval that could either be based on the similarity of the topic with video transcript or video captions. This is because the answer could be either in visual or audio modality or jointly in both for questions based on topic text. Therefore, we use  $sim_t(c, v)$, $sim_f(c, v)$, and $sim_v(c, v)$ for the transcript, video, and joint similarity computations, respectively. For transcript similarity-based retrieval, we define

\begin{equation}
   sim_t(c, v) = \langle E_C(c)^T,  E_T(t)\rangle  
   \label{eqn:similarity}
\end{equation}

where $E_C$ is the topic (context) representation, and $E_T(t)$ is transcript representation. Similarly, for video-frame similarity-based retrieval, we define

\begin{equation}
   sim_f(c, v) = \langle E_C(c)^T,  E_T(f)\rangle  
   \label{eqn:similarity2}
\end{equation}

where $E_T(f)$ is video frames representation. Likewise, for similarity comprised of both modalities, we use:

\begin{equation}
   sim_v(c, v) = \langle E_C(c)^T,  [E_T(f) + E_T(t)]\rangle  
   \label{eqn:similarity3}
\end{equation}

where $E_T(f) + E_T(t)$ indicated concatenation of encodings of two modalities.

During training, the encoders are trained so that the similarity function becomes a good ranking function, which is possible by encoding the topic and video representations of correct pairs closer to the encoding space than the irrelevant ones. We describe training steps in the experiments section and provide additional details in SI.

During inference, we encode all videos and index them. Given a topic $c$ at inference, we first retrieve the top $k$ videos with representation closest to $E_C(c)$. For these top videos, we then use an RC model to generate answers to the questions (as described next) relevant to the topic $c$.

\subsection{Reading Comprehension Model for Generating Answers}
The second part of the proposed approach uses a reading comprehension (RC) model. Recent approaches to RC models use the BERT (Bidirectional Encoder Representations from Transformers) architecture to perform reading comprehension tasks. These models are trained on large amounts of text data and can be used to answer questions based on a given context.

The input to a BERT-based reading comprehension model typically consists of two parts: the context and the question (see Fig. ~\ref{fig:rc_model}). The context is a piece of text that provides the background information for the question, while the question is a short piece of text that asks something specific about the context. Both the context and the question are  encoded as sequences of tokens (words or subwords) using a pre-defined vocabulary.

The output of a BERT-based reading comprehension model is a predicted answer to the question. The model first processes the input context and question to create a joint representation of both, then uses this representation to generate a probability distribution over all possible answers. The final predicted answer is the answer with the highest probability in the distribution.

Popular question-answering models like the RoBERTa model \cite{liu2019roberta } can process input token lengths up to 512, but in our case, most video transcripts are longer than 512 words. When the document becomes longer, self-attention's memory and computational requirements grow quadratic. Therefore, long documents remain a shortcoming of attention-based RC models. 

To address the problem of processing longer documents, the Longformer model was introduced in \cite{beltagy2020longformer}, which can process eight times longer than BERT-like models (i.e., 4,096-word tokens). More importantly, in terms of computation, even with longer documents, the Longformer model computation needs to grow linearly with document length. This is possible by combining a windowed local-context self-attention and an end-task motivated global attention~\cite{beltagy2020longformer}. 

We extend the LongFormer model by including a context input that is a joint representation of video transcript text and frame captions. Similar to the BERT Question Answering model, we simultaneously pass the question and context to the Longformer QA model. Though we use the extension of LongFormer as the main model, we compare other RC models in the experiments section. We describe the different parts of LongFormer RC models with multi-modal input in detail here:

\begin{figure}[htb]
    \centering
    \includegraphics[width=0.5\textwidth]{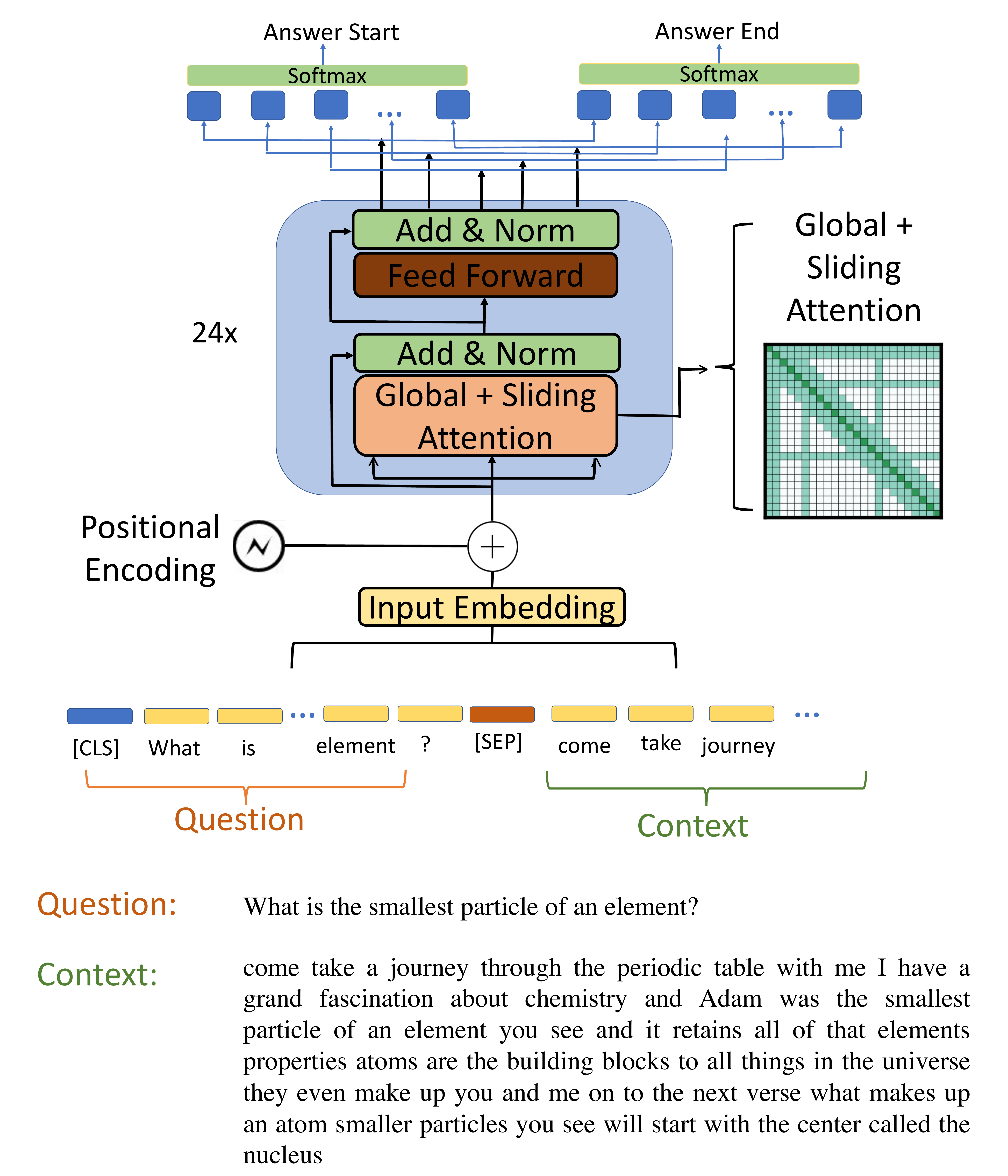}
    \caption[RC Model]{Reading comprehension model with global and sliding attention windows.}
    \label{fig:rc_model}
\end{figure}

\subsubsection{Input Embedding Block}

The Longformer model has an input layer, 12 layers of attention blocks, and an output layer. The input block takes word embeddings, position embeddings, and token-type embeddings. The input context and the query textual representations are fed to the model in the same sequence separated by a special token (`[SEP]' ). This input block embeds all words in the context and in the query into vectors and passes it to the attention block along with the attention mask as shown in Fig. \ref{fig:rc_model}.


%

\subsubsection{Attention Block}
The attention mechanism is a key component of many reading comprehension models, including BERT-based models. The purpose of the attention mechanism is to allow the model to selectively attend to relevant parts of the context when predicting the answer. The attention mechanism works by computing a set of attention weights for each position in the context based on its relevance to the question. In the Longformer model, 12 layers of Transformer Encoders with global and sliding window attention are stacked vertically for building the architecture of the model. Each attention layer comprises attention weights for key $K$, value $V$and query $Q$ items followed by a dense output layer\cite{vaswani2017attention}. Given liner projections $Q, K, V$, a transformer model computes attention score as:

\begin{equation}
    Attention(Q, K, V) = softmax(\frac{QK^T}{d^{\frac{1}{2}}_k}) V
\end{equation}

To better handle longer documents, the Longformer model proposed two sets of projections, $Q_s, K_s, V_s$ (sliding window), and  $Q_g, K_g, V_g$ (global), representing sliding window attention and global attention, respectively. This allows Longformer to use different types of attention, which are less computationally expensive and improves performance on downstream tasks using longer documents.

\begin{equation}
    Attention(Q_s, K_s, V_s) = softmax(\frac{Q_sK_s^T}{d^{\frac{1}{2}}_k}) V_s
\end{equation}

\subsubsection{Output Block}
We add a question-answering head (output block) on top of the model to perform the question-answering task. This question-answering head aims to find an answer's start and end index for a given context. The output block predicts an answer span by computing the scores of each token, the start token, and the end token. The output of the RC model returns two sets of vectors \emph{Start Logits} and \emph{End Logits} and uses the $softmax$ function of these two vectors, which returns indices of the start and end positions of the answer predicted by the RC model.

If $P_i$ is the Longformer representation for the $i-th$ context to the RC model, we find the answer's start $P_{start,i}$ and the end $P_{end,i}$, by formulating the problem as:

\begin{equation}
    P_{start,i}(s) = softmax(P_iW_{start})_s
\end{equation}

\begin{equation}
    P_{end,i}(t) = softmax(P_iW_{end})_t
\end{equation}



where $P_i \in R^{Lxh}$ and  $W_{start}, W_{end} \in R^h$ are vectors that are learned during pre-training, where $L$ is the maximum length of the passage and $h$  the hidden dimension size.


\subsection{Multiple-Choice Answer Extraction Model}
A reading comprehension model generates a textual response by mapping start and end tokens to the input context. However, children's textbook chapters are composed of multiple-choice questions and answers, so the textual answers must be converted to one of the multiple choices for final evaluation. This conversion process involves comparing the output of the RC model to the text in each multiple-choice option to select the closest choice.

Previous research has used two main methods for selecting a choice from the multiple-choice options. The first approach involves using a shallow neural network classifier, while the second approach uses similarity in the embedding space of the predicted answer and the multiple choices to identify the most similar option. The final selected answer is compared against the ground truth answer to determine whether the RC model's answer is correct or incorrect. Both approaches that we use for this work are further explained in the following section:




        



\subsubsection{Neural Network for Multiple Choice (NNMC)}
One way to select the closest choice from the multiple-choice options is to use a neural network with a dense layer and a softmax layer. This approach relies on a pre-trained model trained on the SQUAD dataset. Additional details about the pre-trained models can be found in the supplemental information section.

\subsubsection{Closest Language Embedding Model (CLEM)}
Our proposed method, called the Closest Language Embedding Model (CLEM) calculates the similarity between the answer predicted by the RC model and the multiple-choice options using \emph{fasttext}\cite{joulin2017bag} text embedding. We tune \emph{fasttext} language model on video transcript data to contextualize the embeddings. To select the most similar option from the multiple choices as the answer to the question, we use

\begin{equation*}
    \hat {c} = \operatorname*{argmax}_{c_i} \big(| Cosine(fasttext(c_i) - fasttext(ans) |\big)\\
    \label{eqn:similarity}
\end{equation*}

 where, $c_i$ represents the answer choices, $ans$ represents the answer generated by the RC model, $fasttext$ is text encodings generated using fasttext model and $ \hat c$ is the final selected option. 

 





\section{Dataset}\label{sec:dataset}
We use three datasets in this research. The first dataset is a new dataset of videos from the top YouTube Kids (YTK) channels. The second dataset is another new video dataset that has manually labeled questions and answers based on 100 videos for YouTube. The third dataset is a textbook question answers dataset that has been obtained from prior research. We describe all datasets in detail next.

\subsection{YouTube Videos Dataset}

We created our first dataset by collecting 85,976 videos from the top YTK 100 channels. Using the Google API, we obtained YouTube data and a list of the top 100 "made for kids" channels from the SocialBlade website that ranks "made for kids" channels\footnote{https://socialblade.com/youtube/top/category/made-for-kids}. We then used the Google API to retrieve the list of all videos from these top 100 channels, resulting in a total of 85,976 videos. We also used the Google API to gather video descriptions and transcripts for these videos. Additionally, we obtained copies of publicly available videos using the YouTube-dl library, which has been used in previous research on YouTube videos. We used these videos to obtain video frames that were evenly sampled per minute, with as many video frames as the duration of the video in minutes. Because not all listed videos on YouTube have publicly available transcripts and video files, we only included videos that had both of them available in our final dataset.

The final video dataset is summarized in SI, along with information on the channel name, the total number of videos we were able to retrieve for the channel, and the total number of views for all videos in the dataset. It's important to note that these view counts only consider the videos that are in the dataset, i.e., not necessarily all videos of the channel. As we can find, the ``Jugnu Kids'' channel has the largest number of videos (4676) in the dataset, and the "Cocomelon" channel has the highest number of views (152 Billion) among the channels in the dataset.

\subsection{Children's' Videos Question Answering (\texttt{CVQA}) Dataset}
While RC models have been used with text- and image-based input, their usage with video data is limited. To validate if RC models could be used with video data for answering factual questions, we need a dataset with  questions that can be answered by seeing the videos. As no existing data sets on children's videos are labeled for questions and answers, we create a small dataset of approximately one hundred videos, manually labeled for 653 questions that can be answered using the information in the videos. The labeling was conducted by Research Assistants, who are well-versed in English, and a faculty member associated with a large private university who earned his Ph.D. in the USA. Unlike many classification tasks (e.g., sentiment classification) where the labelers could have differing opinions leading to different labels, extracting factual questions present in videos does not have that concern. What is possible, though, is that a labeler might miss some questions that could be considered legit questions by another labeler. This limitation is acceptable for our first experiment that uses this data, as the goal is not to evaluate the recall of questions but  to find if RC models could answer questions that are indeed discussed in the videos. Therefore, multiple labelers did not label the same video.

\begin{table}[ht!]
\scriptsize
\caption{\texttt{CVQA} Dataset includes ground truth questions and answers from Children's YouTube Videos (along with the proposed model's sample outputs)}
\centering
\begin{tabular}{|p{2.0cm}|p{1cm}|p{1.2cm}|p{1.5cm}|p{1.0cm}|}
 \hline
Question & Ground Truth & Model Output & YouTube Video Id & Time Stamp\\ [0.5ex]
 \hline
 \hline
What is at the center of an atom? & Nucleus & the nucleus & QP0uqR7A1WQ & 00:35\\\hline
In which place, are no native ants present? & Greenland &  in Greenland & Fw1\_1\_yKZ78 & 01:24\\\hline
What do the teeth play a key role in? & Digestive System &  digestion & XdkOQJTei6c & 00:43\\\hline
How many gallons of water does a horse need to take every day? & 5-10 gallons &  5 to 10 gallons & cPiW2VFz37g & 02:00\\\hline
 \end{tabular}
 \label{tab:labeled_dataset}
\end{table}

We show some examples of questions and answers in Table \ref{tab:labeled_dataset} along with the video's unique id and approximate time-stamp around which the information to answer questions was presented in the video. As one can observe, the questions are factual (to mimic questions in the \texttt{TQA} dataset), and answers are typically one to three words. More details on video labeling are provided in the SI.

\subsection{Textbook Question Answering (\texttt{TQA}) Dataset}

We aim to evaluate the quality of academic videos for children by comparing them to textbook questions. One dataset that we use for this purpose is the TextBook Question Answering (\texttt{TQA}) dataset proposed by \cite{kembhavi2017you}, which is composed of questions and answers on topics from Science textbooks\footnote{https://allenai.org/data/tqa}. This dataset was created from the ck12.org website\footnote{http://www.ck12.org/} and includes 1,076 lessons, with each lesson covering multiple topics. For example, a lesson on Earth Sciences may include topics on Geology, Oceanography, Astronomy, Climatology and Meteorology, and Environmental Science. Each of these lessons has related questions that can be answered using the topic context. 

The questions are on a range of topics including \textit{earth science and its branches, renewable resources and alternative energy sources, human digestive system, etc.,}, and  each of these topics contains multiple sub-topics. Overall, the dataset has 26,260 questions and their answers. Most questions are multiple-choice questions, with two to four answers. In SI, we provide additional details about this dataset, visualizing the dataset structure, and sample questions and answers.

\section{Experiments and Results}
\label{sec:exp_results}
Our experimental design is divided into three parts. The first part validates our proposed approach of using an RC model for question answering using questions extracted from children's videos. In this part, we also compare various RC models, showing how different RC models compare for varying video lengths and the complexity of questions. The second part estimates the academic quality of channels using questions from children's textbooks (\texttt{\texttt{TQA}} dataset). Finally, the last part shows what academic topics are covered by videos and compares different channels for their academic quality. For each part, we discuss the implementation details followed by results.

\subsection{Experiment 1: Evaluating RC Models on \texttt{CVQA} Question Answers Dataset}
This experiment aims to validate that RC models can indeed be used for answering questions from YouTube videos. We also want to compare the state-of-the-art RC models so that the best RC models can be picked for the next set of experiments. To evaluate the RC models on the \texttt{CVQA} dataset, we use four RC models. Our analysis pipeline expects video data and a question-answer set, as in the \texttt{CVQA} dataset discussed earlier. The RC model reads and comprehends the video data (transcript, caption, transcript + caption) to answer questions so that the contribution of each modality could be identified independently. Based on the number of questions from \texttt{CVQA}, correctly answered by the RC model using the videos data, we estimate and compare the performance of different RC models (see Tbl. \ref{tbl:rc_eval}), performance on videos with different duration (see Fig. \ref{fig:passage_len}), and different video modalities (see Tbl. \ref{tbl:rc_eval2}). 

\subsubsection{Training and Evaluation}
For evaluating RC models on \texttt{CVQA} dataset, we use pre-trained RC models, so no additional training is needed. More information on pre-trained models is available in the SI.  We evaluate the RC model output with the ground truth answer. As an RC model's output could be more than a few words, we use 1) stemming and 2) word matching to find if all words  in the ground truth answer are present in the RC model output. If all words in the ground truth answer are present in the output text of the RC model, we mark the prediction as correct; otherwise, we mark the prediction as wrong. We use the correct and incorrect labels for each question to calculate the accuracy, precision, and F1-Score for all models, as shown in Tbl. \ref{tbl:rc_eval}.


\begin{table}[htb!]
\scriptsize
\caption{Models performance on  \texttt{CVQA} questions from YouTube Videos. * implies the model restricts the input to 512 characters.}
\centering
\begin{tabular}{|p{2.7cm}|p{1cm}|p{1cm}|p{1cm}|}
 \hline
 Model &  Accuracy &  Precision  & F1-Score \\ [0.5ex]
 \hline
 \hline
Longformer & \textbf{0.79} & \textbf{0.79}  & \textbf{0.88}\\\hline
BiDAF & 0.52 & 0.52  & 0.68 \\\hline
BiDAF-ELMo & 0.59 & 0.59  & 0.74 \\\hline
Transformer QA* & 0.42 & 0.2  & 0.59 \\\hline
 \end{tabular}
 \label{tbl:rc_eval}
\end{table}


\subsubsection{Results}
\begin{figure}[htb!]
    \includegraphics[scale=0.35]{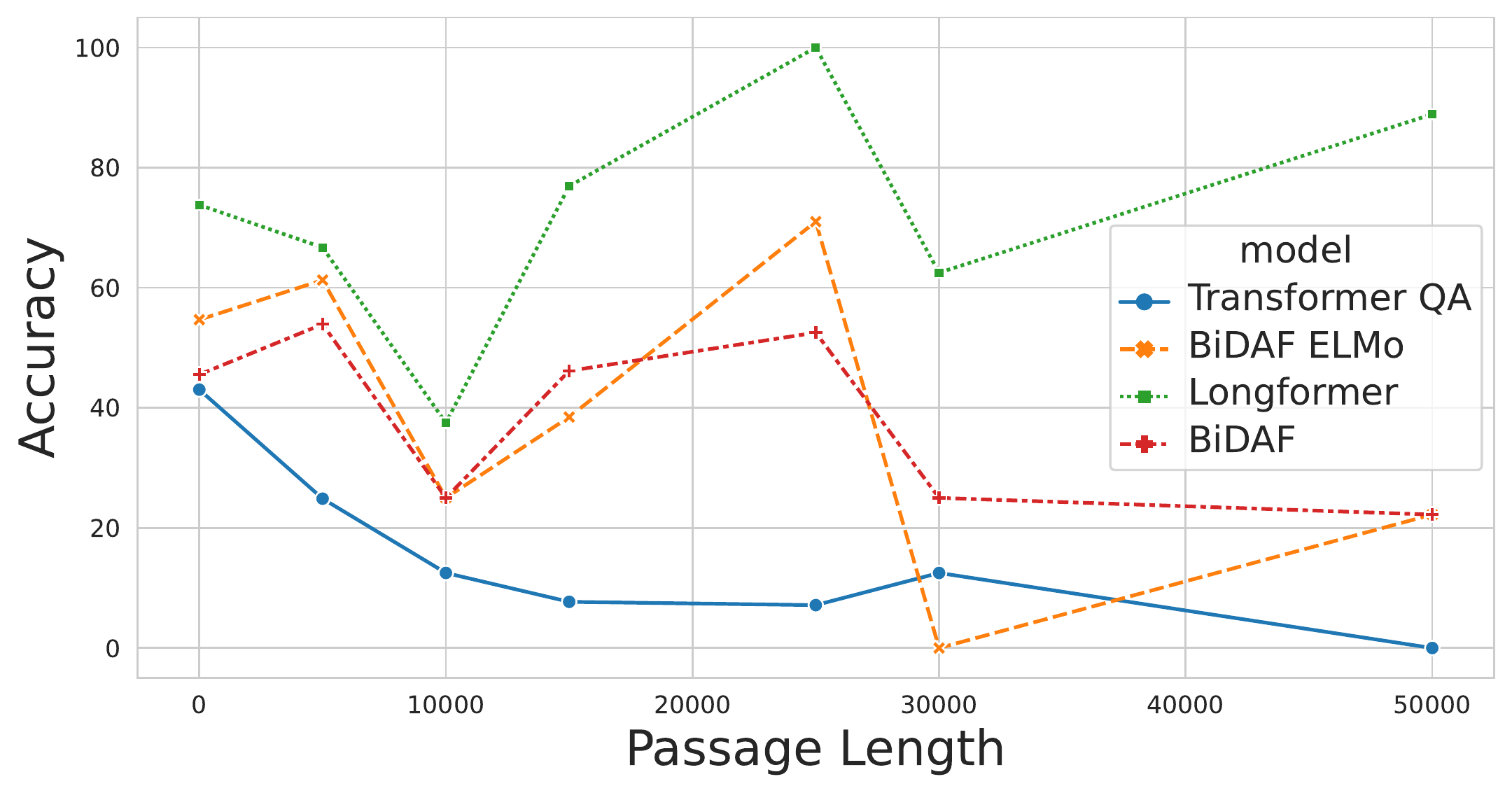}
    \caption[RC Model]{Comparing RC models of varying passage length}
    \label{fig:passage_len}
\end{figure}

We present the results for the RC model in Tbl. \ref{tbl:rc_eval}. As one can observe in the table, the `Longformer' model achieves the highest F1-score of 0.88, followed by `BiDAF-ELMO'. The Longformer model, proposed to be used because the videos could be long, is more accurate and precise, leading to the highest F1 score. The other models have a substantially lower performance, likely because the longer video transcripts are unsuitable for BiDAF and Transformer QA models (see Fig.~\ref{fig:passage_len}). 


We further evaluate the models on different parameters. For this analysis, the length of the input text is a major consideration, as the length of videos varies a lot. We present the results for the same in Fig. \ref{fig:passage_len}. For creating the figure, we group all transcripts in a passage length in multiplications of 5000 (i.e., 0-5000, 5000-1000, ..), showing the mean accuracy for the passages in that group. We find that the Longformer model consistently performs better than other models and improves as the passage length increases. The Bidaf and Bidaf-Elmo models perform well for shorter text, but the accuracy decreases with longer text, especially after the passage length crosses 30,000 limits. For the longest text of 50,000 words, BiDAF models have a final accuracy of over 20\%. On the contrary, the accuracy of the Longformer model remains consistent and above 85\% for very long passages.


\subsubsection{Ablation Study}
We study how the RC models use different features. Our input data comprises video transcripts and captions; therefore, the most obvious question is which of the two modalities is more critical. For this analysis, we look at the Longformer model, the best model we have found for the overall video data, and look for the relative benefits of  the different modalities. We also explore how the model performs for simple and complex questions, an important aspect of evaluating RC models.


\begin{table}[h!]
\scriptsize
\caption{Performance of Longformer model on different types of questions}
\centering
\begin{tabular}{|p{1.4cm}|p{1.8cm}|p{0.8cm}|p{0.8cm}|p{1cm}|}
 \hline
 Question Type & Context & Accuracy & Precision & F1-Score \\ [0.5ex]
 \hline
 \hline
All Questions & Transcript & 0.79 & 0.79 &  0.88 \\ \hline
All Questions & Caption & 0.72 & 0.72 &  0.83 \\ \hline
\hline
Complex & Transcript & 0.74 & 0.74 & 0.85 \\ \hline
Complex & Caption & 0.66 & 0.66  & 0.8 \\ \hline
\hline
Simple & Transcript & 0.8 & 0.8  & 0.88 \\ \hline
Simple & Caption & 0.73 & 0.73  & 0.84 \\ \hline
 \end{tabular}
 \label{tbl:rc_eval2}
\end{table}

In Tbl. \ref{tbl:rc_eval2}, we show the relative feature importance for questions segregated as `simple' and `complex' and for different modalities. In our dataset, simple questions have answers in a single sentence, and complex questions have answers that need to combine the information in multiple sentences. As we can observe, the model performs slightly better on simple questions. Moreover, in general, the model is better at answering questions based on transcripts (language modality) than questions based on visual understanding.


\subsubsection{Error analysis}
While the Longformer model can get good accuracy, analyzing cases in which it fails could give valuable insights. We manually looked at some questions on which the model fails and observed two patterns: 1) It fails more often for complex questions, and 2) It fails for many video captions, as video captions do not often encode all information contained in a video frame.  These findings are aligned with the empirical evidence in Tbl.~\ref{tbl:rc_eval2}.

\begin{figure}[htb!]
    \centering
    \includegraphics[scale=0.28]{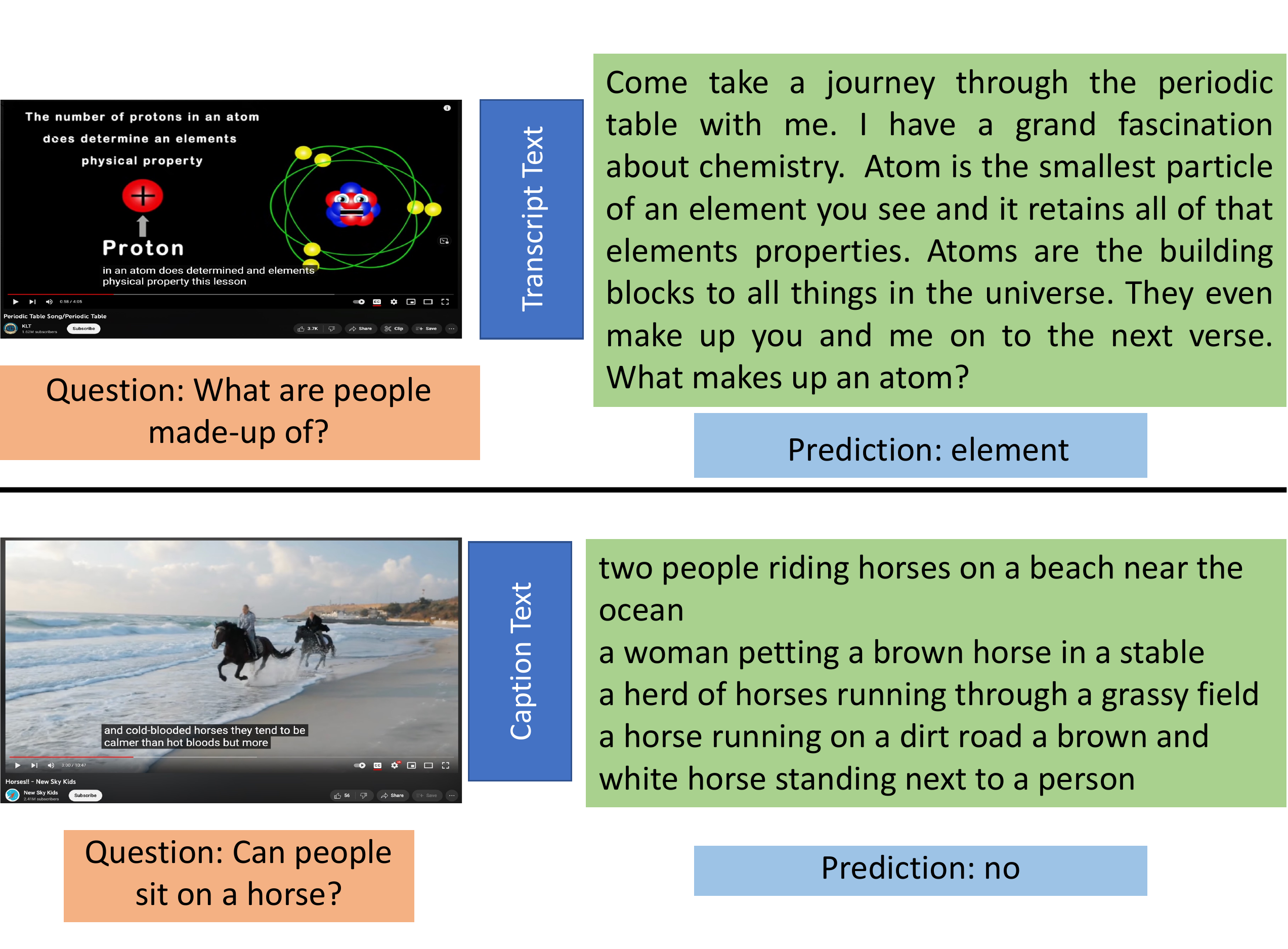}
    \caption[RC Model]{Two example predictions (one for the language modality and one for the visual modality) from the Longformer model when the answer predictions are incorrect. }
    \label{fig:error_analysis}
\end{figure}

\begin{table*}[htb!]
\caption{Accuracy of Reading Comprehension (RC) models with questions from \texttt{TQA} dataset and transcript and captions obtained from YTK videos. }
\label{tab.results}
\tiny
\begin{tabular}{|p{3.6cm}|p{1cm}|p{1cm}|p{1.2cm}|p{1.2cm}|p{1.2cm}|p{1.2cm}|p{1.2cm}|p{1.2cm}|p{1.2cm}|}
\hline
Retriever + RC + Answer Extraction Models & Retrieval Measure &Training Dataset & Baseline with Random Text & Transcript (K=1)&  Transcript (K=3)& Caption (K=1)& Caption (K=3) & Transcript + Caption (K=1)& Transcript + Caption (K=3) \\\hline
\hline
\multicolumn{10}{|l|}{\textbf{\textit{Baseline Models}}} \\\hline

Dense Transcript Retriever + Longformer + NNMC  & Transcript & \texttt{SQuAD}v1 & 32.01 & 19.11 & 39.08 & 23.86 & 36.54 & 20.58 & 38.82\\\hline
Dense Transcript Retriever + Longformer  + CLEM & Transcript &\texttt{SQuAD}v1 & 33.19 & 21.79 & 73.36 & 22.30 & 66.55 & 19.93 & 71.06\\\hline

Dense Caption Retriever + Longformer + NNMC & Caption & \texttt{SQuAD}v1 & 32.01 & 18.20 & 23.53 & 15.39 & 23.12 & 20.74 & 23.56\\\hline
Dense Caption Retriever + Longformer  + CLEM & Caption & \texttt{SQuAD}v1 & 33.19 & 21.37 & 23.14 & 22.65 & 22.76 & 22.13 & 22.53 \\\hline
\hline






\multicolumn{10}{|l|}{\textbf{\textit{Proposed Models based on Longformer}}} \\ \hline
Dense Transcript Retriever + Longformer + NNMC & Transcript & \texttt{SQuAD}v1 & 32.01 & 53.28 & 45.41 & 24.14 & 37.76 & 36.27 & 54.69 \\\hline
Dense Transcript Retriever +  Longformer + CLEM & Transcript & \texttt{SQuAD}v1 & 33.19 & 67.34 & 76.02 & 49.60& \textbf{69.06} & 55.69 & 76.27 \\\hline 

Dense Caption Retriever  + Longformer + NNMC & Caption & \texttt{SQuAD}v1 & 32.01 & 30.65 & 47.11 & 16.77 & 34.79 & 30.93 & 30.94 \\\hline
Dense Caption Retriever  +  Longformer + CLEM & Caption & \texttt{SQuAD}v1 & 33.19 & 39.41 & 67.60 & 32.64 & 46.72 & 35.61 & 43.74\\\hline
\hline

\multicolumn{10}{|l|}{\textbf{\textit{Proposed Models based on Longformer with Mult-modal Retriever}}} \\ \hline
Multi-modal Retriever + Longformer + NNMC & Transcript + Captiopn & \texttt{SQuAD}v1 & 32.01 & 50 & 53.83 & 30.16 & 33.74 & 47.79 & 51.37 \\\hline
Multi-modal Retriever +  Longformer + CLEM & Transcript + Captiopn & \texttt{SQuAD}v1  & 33.19 & \textbf{71.48} & \textbf{76.44}& \textbf{52.30} & 56.48  & \textbf{72.89} & \textbf{78.09} \\\hline

\end{tabular}
\end{table*}

In Fig. \ref{fig:error_analysis}, we show two examples where the model  made incorrect predictions. The top sub-figure shows a prediction when passing the video transcript text, and the bottom sub-figure uses video-frame captions. In the first case, the incorrect prediction is because the transcript text does not include `people' as a word in the text, and the model could not connect `you and me' to people. In the bottom example, the error is likely because the model could not connect `sitting' as a part of riding.

To summarize the results of experiment 1, we find that different models perform differently when it comes to analyzing YouTube video data. The best model, that is, the Longformer, beats the second-best model BiDAF-ELMO by 18\%. The length of the input string is the major determinant in the performance of the models, and LongFormer, because of its capability to handle longer texts, is better positioned for analyzing the videos. 

\subsection{Experiment 2: Quantifying the Academic Quality of Videos using \texttt{TQA} dataset}
In this experiment, we use the \texttt{TQA} dataset to estimate the academic quality of videos and channels. As mentioned earlier, the inference on RC models is a time taking process, and even with  the state-of-the-art models, reading all video data one by one and answering all questions in the \texttt{TQA} dataset would take a prohibitively long amount of time. Therefore, generating video encoding and retrieving $k$ top videos for each topic in the \texttt{TQA} dataset is preferred for this analysis. 

\subsubsection{Training and Evaluation}
To train the video retriever model, we start with pre-defined weights from the dense encoder network architecture as used in Karpukhin \textit{et al.}~\cite{karpukhin2020dense} and then fine-tune the model weights on our dataset. The finetuning of a model requires the ground truth of positive and negative videos. However, there is no ground truth for training in our context, as YouTube videos are not mapped to \texttt{TQA} questions. To address this, we create pseudo-positive and negative videos using a BM25 retriever model \cite{robertson2009probabilistic} based on transcript or caption or both transcript and caption data. The videos that lead to correct answers are grouped as positives, while those that lead to incorrect answers are grouped as negatives. We use these negative and positive videos to define a contrastive loss function to train the encoder models (more details in SI).

We tune three separate models named `Dense Transcript Retreiver', `Dense Caption Retreiver', and  `Multi-modal Retriever' based on the data used for training, i.e., transcript, caption, or joint transcript and caption, respectively. During inference, we use the encoded video data and first retrieve the top $k$ ($k$=1 or 3 in experiments) videos with representation closest to the learning topic description (from the \texttt{TQA} dataset) based on retrieval measure (transcript, caption, or joint). For these selected videos, we then use an RC model to generate answers to the questions. After retrieval, we use the Longformer RC model in the second stage with pre-trained weights. The Longformer RC model  is pre-trained on the Stanford Question Answering Dataset (\texttt{SQuAD}) for question answering. More details on the RC pre-trained models are in SI. We finally use the multiple-choice answer extraction models (either NNMC or CLEM) to measure the accuracy of question answering based on video data.


\subsubsection{Results}
As mentioned earlier, given a \texttt{TQA} topic, we use video retrievers to retrieve the top $k$ videos. We then use the LongFormer model, the best model we found in our earlier experiments, to answer the topic questions and determine how many questions are correctly answered. This results in a dictionary composed of $\{ video: score\}$, where the score indicates the number of questions that could be correctly answered. We use this dictionary to estimate the accuracy in answering questions for different combinations of models (Retriever + RC + Answer Extraction Model) and retrieval measures (transcript, caption, and transcript + caption). We present the summary of this analysis in Tbl. \ref{tab.results}. 

\begin{figure*}[htb!]
    
    \includegraphics[width=0.9\textwidth]{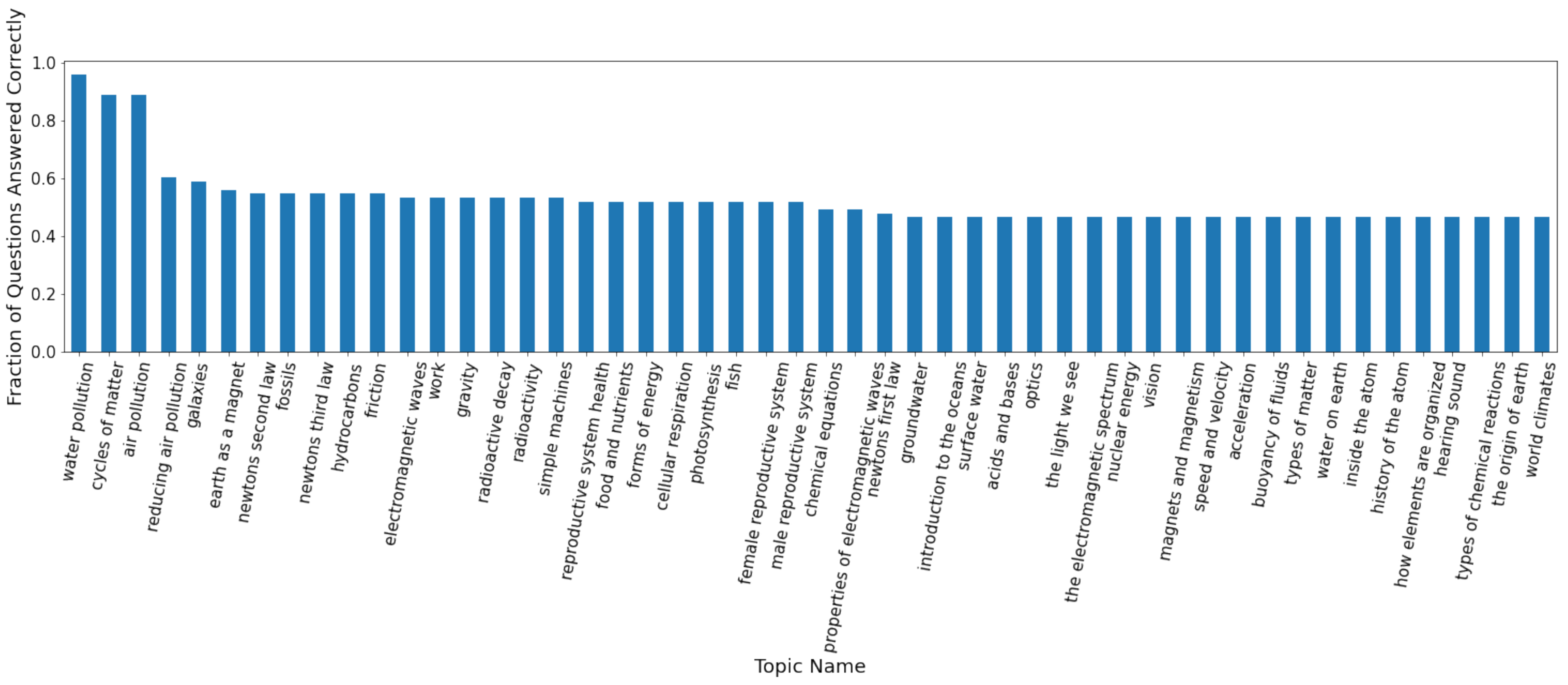}
    \centering
    \captionsetup{justification=centering}
    \caption{The top topics (from children's textbook) covered by the top 100 YouTube channels }
    \label{fig:overall}
\end{figure*}

For a baseline, we use the iconic sentence `Colorless green ideas sleep furiously'~\cite{chomsky2002syntactic} as input to the RC models. This `random text' is not entirely random but has been used in prior work for similar purposes. Given that random text has no helpful information, the RC model output will likely result from its own biases. Therefore, it serves as an excellent baseline to show if videos can be used to answer the questions correctly. The accuracy of models using random text as inputs are below 34\%, slightly better than the random chance of picking one of the four answers in the multiple-choice questions. 


In Tbl. \ref{tab.results}, we show our main results. The table heading refers to the type of model (Retriever + RC + Answer Extraction Model), the retrieval measure (transcript or caption or transcript + caption), the training dataset for the pre-trained RC model, and accuracy scores for the baseline using random text. We also show accuracy when video transcripts, video captions, and joint transcript and captions are used for the top one or three videos ($k= 1$ and $k=3$). In the table, Transcript (column name) implies audio transcription of videos, and Caption (column name) implies caption text generated from video frames are model's input. Transcript + Caption implies both transcripts and captions are passed as input to the RC model.


Our results show that transcript-based retrievers generally perform better than caption-based retrievers. This is not entirely unexpected, as even in Experiment 1, we observed that captions do not capture all useful information in video frames and, therefore,  are not as effective as the transcripts. The difference in the performance is starker in this experiment, likely because the kind of questions in the \texttt{TQA} dataset (mostly factual) is different from the text generated by caption generators (mostly descriptive about the video frames).


However, the multi-modal retriever still performs the best, indicating that the video frames are essential to get the best performance (78.09\%) while a reasonable performance (76.27\%) could already be obtained just by using readily available video-transcript data. This implies that in some settings, the heavy processing needed to use video frames is not justified if the additional gain of 2\% is not essential.  We also observe that a higher value of $k$ leads to better performance, indicating that  three videos are better positioned to answer \texttt{TQA} questions correctly. This is not unexpected given that additional video means more information is available for answering questions. 

In Fig.  \ref{fig:overall}, we have visualized the top topics YouTube videos cover. As one can observe, specific popular topics (e.g., `water pollution', `air pollution' etc.) are well covered and can be used to answer textbook questions correctly. However, the figure does not show the topics for which the platform has fewer videos. This information is helpful for video creators that can focus on creating more novel content that caters to the needs of middle school children.

 To summarize, the multi-modal retriever model that used  transcripts and captions performed best. What is perhaps pleasantly surprising is that YouTube videos can correctly answer around 78\% of the questions in the middle school textbook curriculum. This does not necessarily mean that all YouTube videos are good, but rather videos from the top 100 channels, if picked appropriately, could lead to substantive learning experiences for children. To understand how channels differ, we present a detailed comparative analysis of YTK channels in the next section.

\subsection{Experiment 3: Comparing Different Channels}
\begin{figure}[h!]
    \centering
    \includegraphics[width=0.44\textwidth]{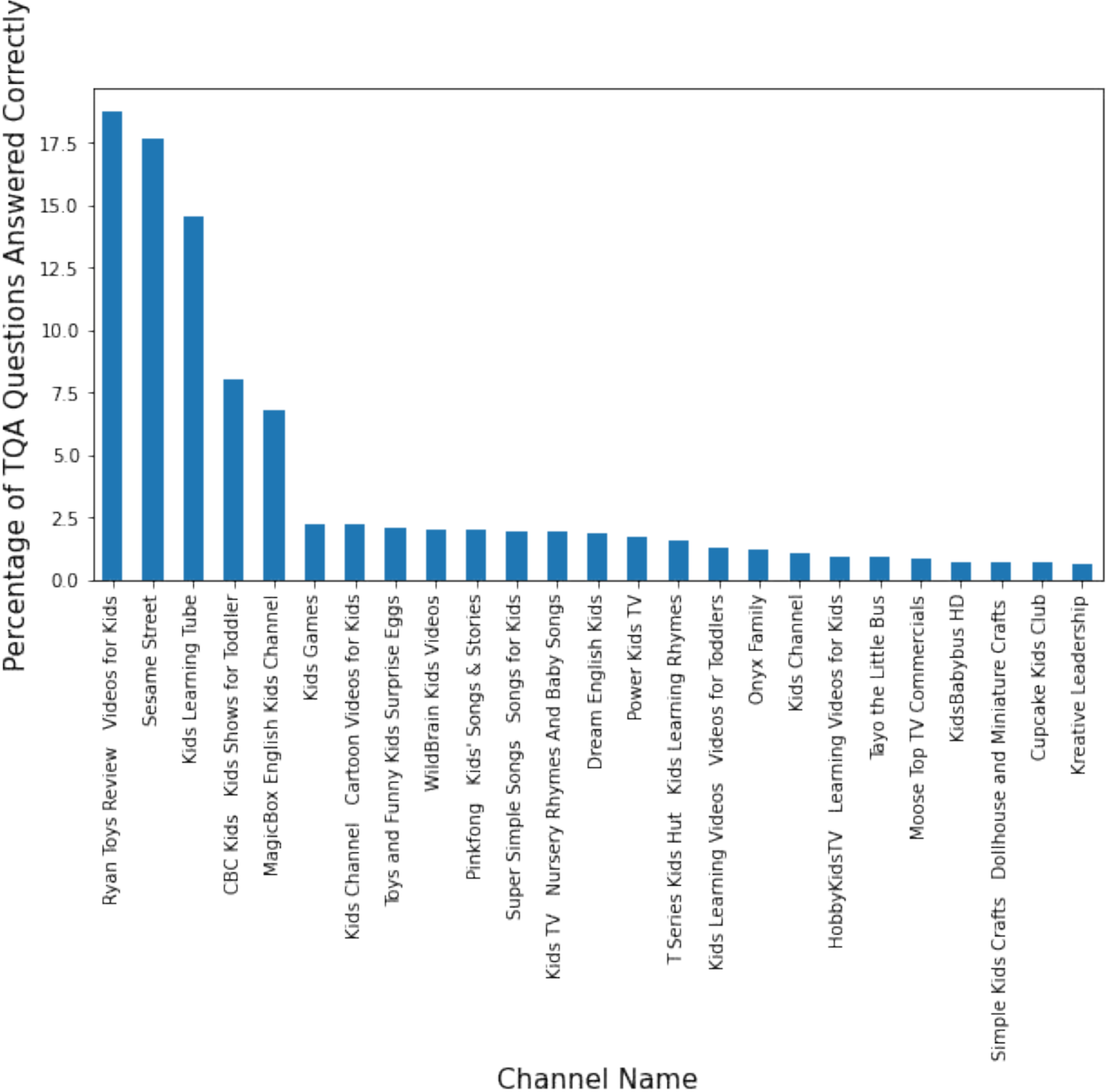}
    \caption[RC Model]{Percentage of \texttt{TQA} questions answered correctly by different channels}
    \label{fig:compare_channels}
\end{figure}

In this section, we compare different channels based on the academic quality of the videos they post. We  used the results $\{video: score\}$ from the last experiment to aggregate the number of questions that can be answered using the videos posted by channels and visualize the aggregate score for the top channels. Figure \ref{fig:compare_channels} shows a bar plot for each channel showing the percentage of \texttt{TQA} questions answered correctly. As one can observe, some channels like `Ryans Toys Review' and `Sesame Street' are doing reasonably well vis-a-vis \texttt{TQA} questions that can be answered, whereas most channels have low scores.

We finally prove a proxy measure for the academic quality of channels. We define the academic quality of a channel $EQ_{ci}$ as the percentage of total \texttt{TQA} questions that can be correctly answered per video by channels. Dividing the number of questions answered correctly by the total number of videos posted by a channel normalizes a high video count.

\begin{equation}
    EQ_{i} = \frac{TQA-questions-answered-by-channel_i}{N_{i}} * 100
\end{equation}

where $N_{i}$ the total number of videos in channel $i$, and $EQ_{i}$ is the academic quality of the channel. For determining the $EQ$ of channel $i$, we chose the model \textit{Multi-modal Transcript Retriever +  Longformer + CLEM} from the last experiment.

\begin{figure}[htb!]
    \centering
    \includegraphics[scale=0.39]{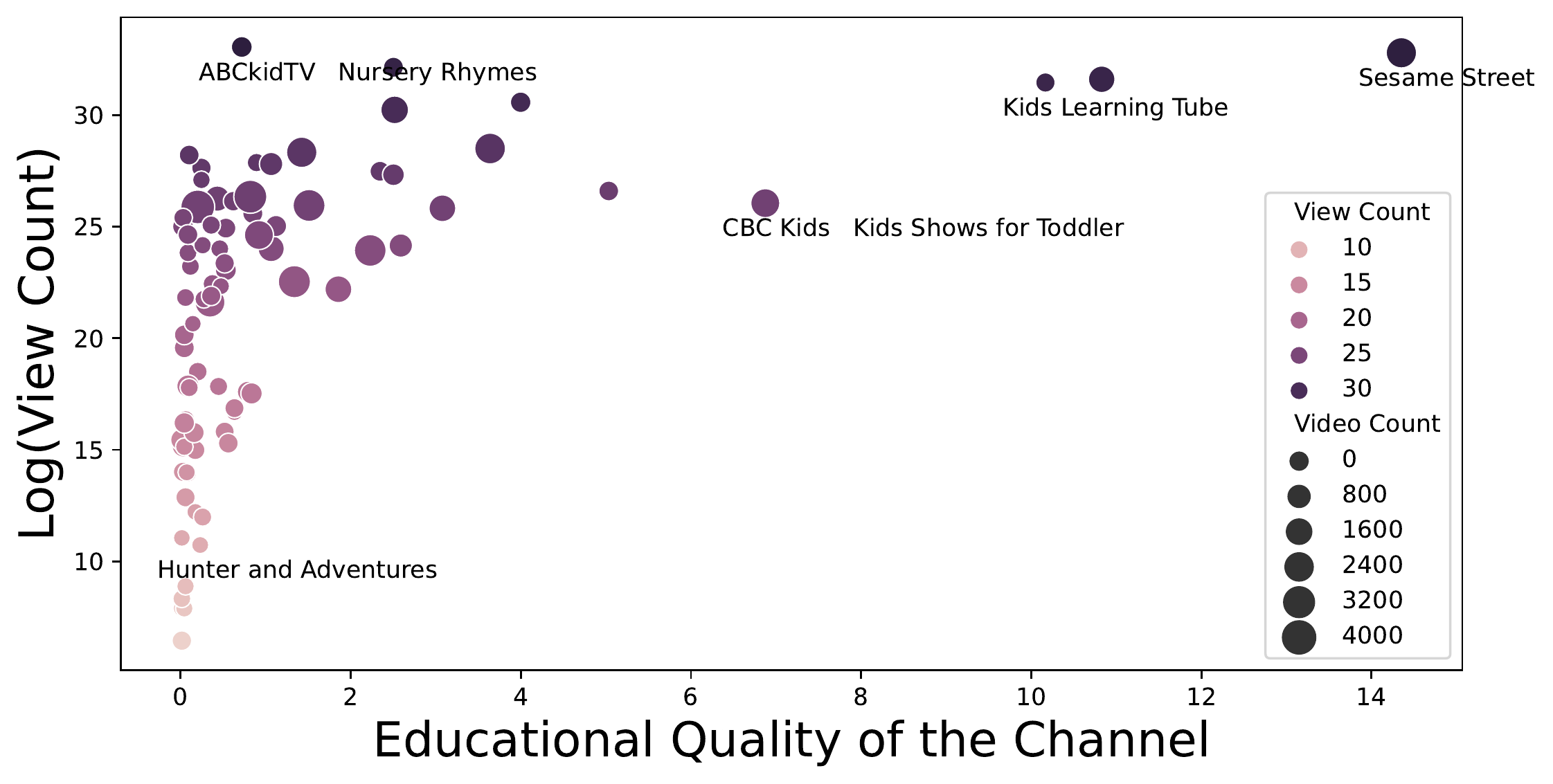}
    \caption[RC Model]{Academic quality of top viewed channels. Each dot represents a YouTube Kids channel, wherein the dot size indicates their relative popularity.}
    \label{fig:channel_comparison_views}
\end{figure}

In Fig. \ref{fig:channel_comparison_views}, we present a comparative analysis of the relationship between video views (a proxy for channel popularity) and academic quality across various channels. Our results indicate that there exists a positive correlation between the popularity of a channel and the academic quality, as evidenced by the higher percentage of \texttt{TQA} questions answered using videos in the channel. This trend can be attributed to the recent emphasis on the quality of content  by YouTube, leading to increased production of high-quality videos by prominent content creators. 

However, it is noteworthy that the majority of channels in our dataset exhibit relatively low levels of academic quality. In particular, some channels with  a substantial number of video views (e.g., `ABCKidTV' and `Nursery Rhymes', and `CBC kids') in our sample also displayed low academic quality. Our findings have important implications for platform owners and content creators, highlighting the need for a continued focus on quality content.







\subsection{Discussion}
With the recent increase in screen time leading to kids spending more time online than with friends, it is important not just for parents but also for video hosting platforms to measure the quality of videos. Our approach not only helps in providing an unbiased and objective measure of academic quality but also enables video hosting platforms to monitor the quality of videos and encourage creators to create high-quality content. 

While our work represents a significant advance, we recognize its limitations. The academic quality of our dataset is limited to middle school textbook standards due to the \texttt{TQA} dataset being drawn from middle school science curricula. We acknowledge this limitation and posit that it is not possible for any dataset to encompass all academic knowledge. Hence, any effort to quantify academic quality must rely on a proxy. Despite these limitations, to the best of our knowledge, we are the first to propose a method for addressing the complex problem of quantifying the academic quality of children's videos.

We also argue that our proposed approach is generalizable. Our context text and the set of input questions came from two different sources, i.e., the YouTube videos and the \texttt{TQA} Dataset. Still, experiment 1 shows that RC models can easily answer questions from \texttt{TQA} using video data. This gives additional validity to the results and also shows the benefits of the proposed approach that allows question answers from a completely different medium to be used in conjunction with YouTube videos.


\section{Conclusion And Future Work}\label{sec:colclusionAndFutureWork}
In this research, we demonstrate the ability of an RC model to assess academic quality by introducing a new dataset consisting of questions and answers from children's videos. We then determine the academic value of the top channels by measuring the number of textbook questions answered correctly by the model. Our analysis of over 80,000 videos posted on the top 100 channels provides a comprehensive evaluation of the academic quality of content on YTK and utilizes a large dataset of middle school textbook questions on various topics. Our findings reveal the academic topics covered in these children's videos, and we compare the quality of the channels.

Our study contributes to the understanding of the academic content quality on YTK and offers valuable insights for enhancing the platform's content selection. Additionally, our research highlights the potential of using RC models to assess the academic quality of online videos, which can be extended to other platforms and domains. Our results can inform the creation of new methods and tools for evaluating the academic value of online videos and promoting high-quality academic content for children.

\bibliographystyle{IEEEtran}
\bibliography{bare_jrnl_compsoc}

%

\begin{IEEEbiography}
  [{\includegraphics[width=1in,height=1in,clip,keepaspectratio]{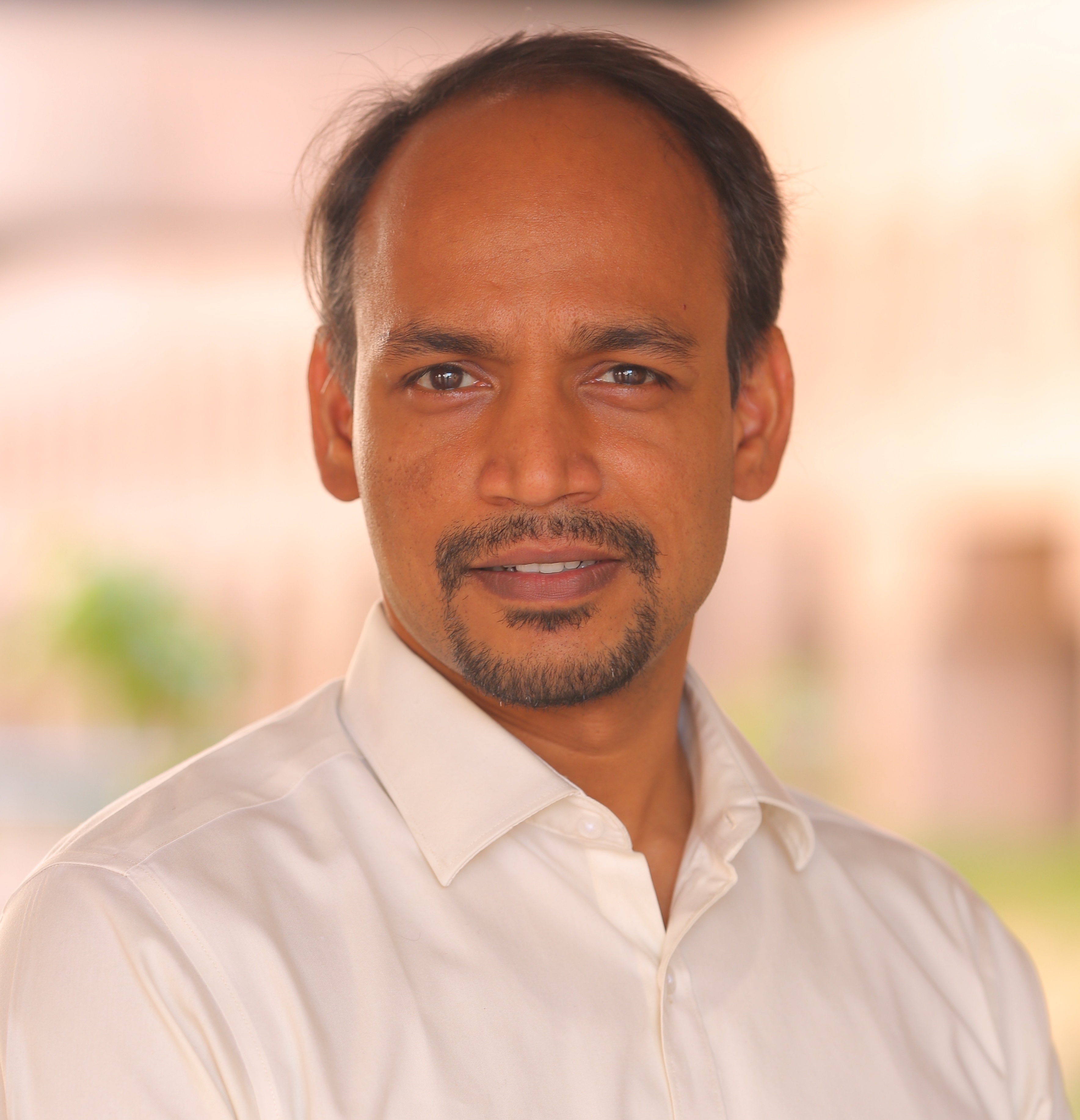}}]{Sumeet Kumar} received his Ph.D. from the School of Computer Science at Carnegie Mellon University. He is currently an assistant professor of Information Systems at the Indian School of Business. He is interested in research problems at the intersection of technology and society. His current focus is on studying issues of concern on children's platforms like YouTube kids.
\end{IEEEbiography}

\begin{IEEEbiography}
[{\includegraphics[width=1in,height=1in,clip,keepaspectratio]{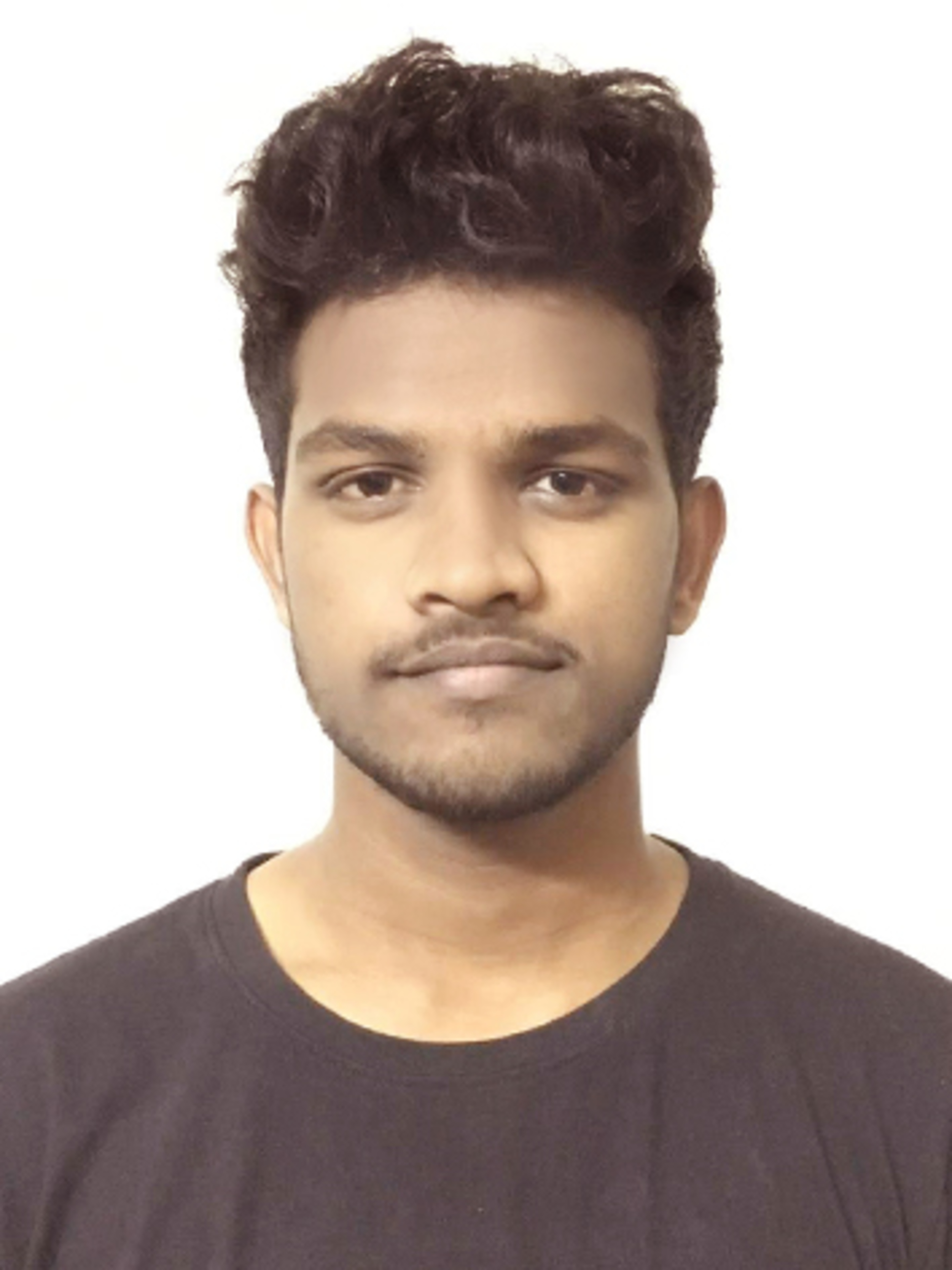}}]
{Mallikarjuna T.}
Mallikarjuna T. is a Research Assistant at the Indian School of Business. He is interested in Deep Learning, Natural Language Processing, and Computer Vision and their applications in the Social and Scientific Domains. His current focus is building ML models for societal-related problems and finding visual biases in YouTube Kids.
\end{IEEEbiography}


\begin{IEEEbiography}
  [{\includegraphics[width=1in,height=1in,clip,keepaspectratio]{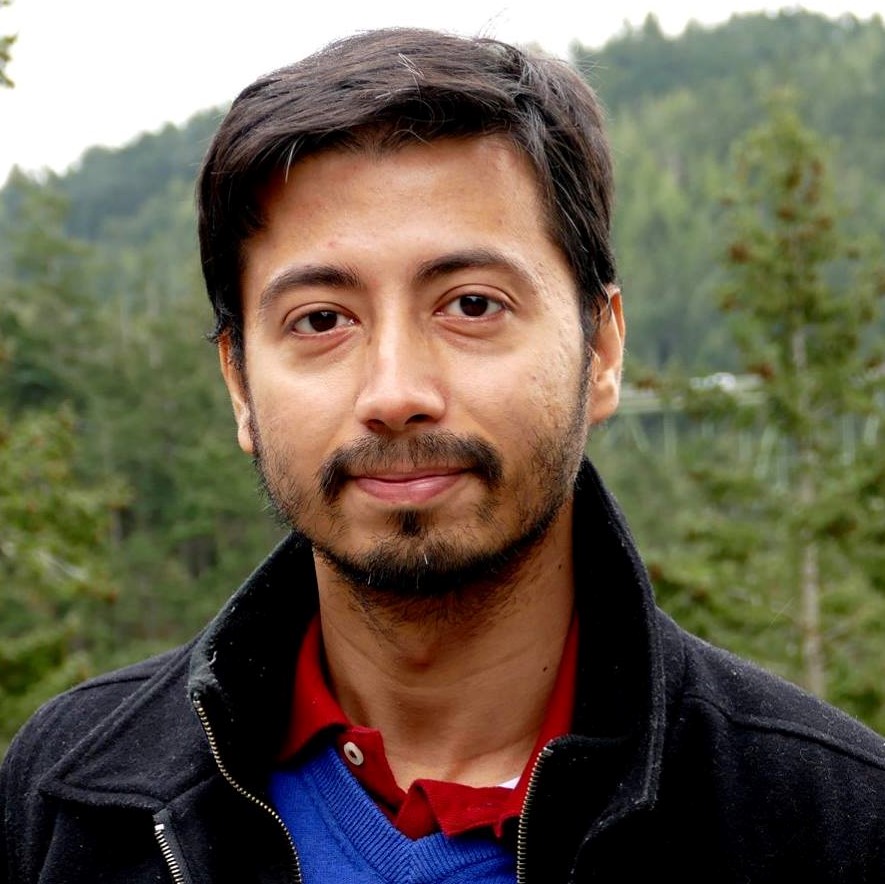}}]{Ashiqur KhudaBukhsh} received his Ph.D. from the School of Computer Science at Carnegie Mellon University. He is currently an assistant professor of Data Science at the Golisano College of Computing and Information Sciences, Rochester Institute of Technology. He is interested in research problems at the intersection of machine learning, natural language processing, and public policy. 
\end{IEEEbiography}




\end{document}